# A Simple Form for the Ground State Rotational Band of even-even Actinide nuclei

**Mohamed E. Kelabi, Khaled A. Mazuz, Eman O. Farhat,
Howida K. Elgowiry, and Samira E. Abushnag**

Department of Physics, Faculty of Science, Al-Fatah University,
Tripoli, LIBYA
P.O. Box 13220, Tripoli, LIBYA

**Abstract**
From Bohr-Mottelson model, a three parametric simple expression for the ground state rotational band of deformed even-even nuclei is deduced by incorporating the variable moment of inertia and the softness parameter. Our obtained results show good agreement with data in comparison with other existing models.

**Introduction**
The ground state rotational band of deformed nuclei was early described by the semi-classical expression [1]

$$E(I) = \frac{\hbar^2}{2\mathcal{J}} I(I+1). \qquad (1)$$

where $\mathcal{J}$ is the nuclear moment of inertia, and $I$ is the nuclear spin state angular momentum, follows the sequence 0, 2, 4, 6,… with even parity. For relatively high spin states, Eq. (1) shows a higher systematic deviation from the energy spectrum [2]. This deviation can be moderated by adding a correction term to Eq. (1), expressive of rotation-vibration interaction [3], [4]

$$E(I) = A\,I(I+1) - B\,I^2(I+1)^2 \qquad (2)$$

where $A\,(=\hbar^2/2\mathcal{J})$ and $B$ need to be determined from experiment[1]. Leaving aside a few very rigid nuclei, Eq (2) shows a noticeable deviation from data and thought as inappropriate to describe rotational energy spectrum [5]. R K Gupta [6] attributed this effect to the variation of moment of inertia $\mathcal{J}$ with nuclear angular momentum $I$, giving

$$E_I = \frac{\hbar^2}{2\mathcal{J}_I} I(I+1). \qquad (3)$$

---

[1] The parameter B is related to the head energies of $\beta$- and $\gamma$-vibrational bands [4].

By using Taylor expansion of $\mathcal{J}_I$ about the ground state value $\mathcal{J}_0$, corresponding to $I = 0$, and incorporating Morinaga's softness parameter [7]

$$\sigma_n = \frac{1}{n!}\frac{1}{\mathcal{J}_0}\frac{\partial^n \mathcal{J}_I}{\partial I^n}\bigg|_{J=0}, \quad n = 1, 2, 3,\ldots \quad (4)$$

Gupta could extract from Eq. (3), by keeping the first and the second order of $\sigma_n$, the following two expressions:

$$E_I = \frac{A\,I(I+1)}{1+\sigma_1 I} \quad (5)$$

$$E_I = \frac{A\,I(I+1)}{1+\sigma_1 I + \sigma_2 I^2} \quad (6)$$

these are known as NS2 and NS3 models, respectively, where the parameters $A$, $\sigma_1$ and $\sigma_2$ can be obtained by fitting with data.

**Approach and Formalism**
In this work, we combine the effect of variation of the moment of inertia $\mathcal{J}$ with nuclear angular momentum $I$, and the effect of nuclear rotation-vibration interaction in a simple expression

$$E(I) = \frac{A\,I(I+1)}{1+\sigma_1 I} - B\,I^2(I+1)^2. \quad (7)$$

We further simplify Eq. (7) by expanding the denominator, using the following form of geometric series [8]

$$\frac{1}{1+\sigma_1 I} = \sum_{n=0}^{\infty}(-\sigma_1 I)^n, \quad |-\sigma_1 I| < 1.$$

The higher powers of $n$ bear less contribution, we therefore keep only the first two terms for $n < 2$,

$$E(I) = A\,I(I+1) - A\sigma_1 I^2(I+1) - B\,I^2(I+1)^2. \quad (8)$$

By setting $a = A\sigma_1$ we obtain an elegant linear expression

$$E(I) = A\,I(I+1) - a\,I^2(I+1) - B\,I^2(I+1)^2 \quad (9)$$

contains three parameters $A$, $a$, and $B$ which can be determined straight forward, using linear method of least squares fitting.

Eq. (9) is our fundamental expression and will be used to calculate the energies of the ground state rotational band of deformed even-even Actinide nuclei.

**Results and Comparison**
In this section we compare our results with other available calculations, namely:
1) The Exponential model (Expo1) [9]

$$E(I) = \frac{\hbar^2}{2\varphi_0} I(I+1) \exp\left[\Delta_0 \left(1 - \frac{I}{I_c}\right)^{1/2}\right]$$

2) The Exponential model (Expo2) [10]

$$E(I) = \frac{\hbar^2}{2\varphi_0} I(I+1) \exp\left[\Delta_0 \left(1 - \frac{I}{I_c}\right)^{1/\nu}\right]$$

3) The Nuclear Softness model (NS3) [6]

$$E_I = \frac{A I(I+1)}{1 + \sigma_1 I + \sigma_2 I^2}$$

4) The Variable Moment of Inertia model (VMI) [11]

$$E_I = \frac{I(I+1)}{2\varphi_I} + \frac{1}{2} C (\varphi_I - \varphi_0)^2$$

In Table 1, we present the result of our calculations tabulated as Linear form of nuclear Rotational and Vibrational interaction (LRV), along with experimental data in comparison other existing models. The corresponding fits of our calculations are given in Table 2. We show in Figure 1, the error of each model relative to experimental data by means of the chi squared per degrees of freedom

$$\chi^2 = \frac{1}{m - p} \sum_m \left(E_m^{exp} - E_m^{calc}\right)^2,$$

where $m$ is the number of data points, and $p$ is the number of free parameters.

**Table 1**. Energy levels of various calculations in the units of [MeV]. Data taken from [12].

|  | $E(I)$ | Data | Expo1 $p = 3$ | Expo2 $p = 4$ | NS3 $p = 3$ | VMI $p = 2$ | LRV $p = 3$ |
|---|---|---|---|---|---|---|---|
| **224Th** | $E(2)$ | 0.0981 | 0.077986 | 0.093478 | 0.096672 | 0.093149 | 0.09013 |
|  | $E(4)$ | 0.2841 | 0.248133 | 0.280255 | 0.28366 | 0.27952 | 0.275322 |
|  | $E(6)$ | 0.5347 | 0.496401 | 0.533632 | 0.534463 | 0.531326 | 0.530558 |
|  | $E(8)$ | 0.8339 | 0.808843 | 0.836207 | 0.834347 | 0.833268 | 0.836662 |
|  | $E(10)$ | 1.1738 | 1.17149 | 1.17734 | 1.17469 | 1.1762 | 1.180301 |
|  | $E(12)$ | 1.5498 | 1.5702 | 1.55136 | 1.55035 | 1.55407 | 1.553987 |
|  | $E(14)$ | 1.9589 | 1.99046 | 1.95645 | 1.95824 | 1.96256 | 1.956072 |
|  | $E(16)$ | 2.398 | 2.41693 | 2.39374 | 2.39668 | 2.39846 | 2.390756 |
|  | $E(18)$ | 2.864 | 2.83288 | 2.86698 | 2.86489 | 2.85922 | 2.868078 |

**Table 1,** continued…

|  | $E(I)$ | Data | Expo1 $p = 3$ | Expo2 $p = 4$ | NS3 $p = 3$ | VMI $p = 2$ | LRV $p = 3$ |
|---|---|---|---|---|---|---|---|
| **226Th** | $E(2)$ | 0.0722 | 0.06488 | 0.072469 | 0.073562 | 0.073053 | 0.071832 |
|  | $E(4)$ | 0.22643 | 0.208936 | 0.226514 | 0.228103 | 0.229274 | 0.225487 |
|  | $E(6)$ | 0.4473 | 0.423293 | 0.446951 | 0.44791 | 0.450497 | 0.446189 |
|  | $E(8)$ | 0.7219 | 0.698886 | 0.72147 | 0.721218 | 0.723647 | 0.721402 |
|  | $E(10)$ | 1.0403 | 1.02638 | 1.04026 | 1.03904 | 1.03992 | 1.040833 |
|  | $E(12)$ | 1.3952 | 1.39605 | 1.39569 | 1.39439 | 1.39317 | 1.396435 |
|  | $E(14)$ | 1.7815 | 1.79756 | 1.7821 | 1.78178 | 1.7789 | 1.782401 |
|  | $E(16)$ | 2.1958 | 2.21964 | 2.19563 | 2.19681 | 2.19369 | 2.195168 |
|  | $E(18)$ | 2.6351 | 2.64958 | 2.6342 | 2.63592 | 2.63486 | 2.633419 |
|  | $E(20)$ | 3.0971 | 3.0721 | 3.0976 | 3.0962 | 3.1002 | 3.098076 |
|  |  |  |  |  |  |  |  |
| **228Th** | $E(2)$ | 0.057759 | 0.056366 | 0.059189 | 0.059217 | 0.058458 | 0.059108 |
|  | $E(4)$ | 0.186823 | 0.182303 | 0.188648 | 0.188624 | 0.188576 | 0.188545 |
|  | $E(6)$ | 0.378179 | 0.370978 | 0.37903 | 0.378884 | 0.380053 | 0.379003 |
|  | $E(8)$ | 0.6225 | 0.615362 | 0.622084 | 0.621846 | 0.623384 | 0.622164 |
|  | $E(10)$ | 0.9118 | 0.9081 | 0.91058 | 0.91038 | 0.91107 | 0.910703 |
|  | $E(12)$ | 1.2394 | 1.24162 | 1.23823 | 1.23825 | 1.2374 | 1.238289 |
|  | $E(14)$ | 1.5995 | 1.60753 | 1.59966 | 1.59998 | 1.59796 | 1.599578 |
|  | $E(16)$ | 1.9881 | 1.99675 | 1.99038 | 1.99076 | 1.98927 | 1.990224 |
|  | $E(18)$ | 2.4079 | 2.39887 | 2.40676 | 2.40639 | 2.40858 | 2.406868 |
|  |  |  |  |  |  |  |  |
| **230Th** | $E(2)$ | 0.0532 | 0.051 | 0.0545 | 0.0543 | 0.05473 | 0.05.4427 |
|  | $E(4)$ | 0.1741 | 0.166771 | 0.1759 | 0.175577 | 0.1782 | 0.175813 |
|  | $E(6)$ | 0.3566 | 0.343282 | 0.3578 | 0.357465 | 0.36276 | 0.357707 |
|  | $E(8)$ | 0.5941 | 0.57629 | 0.594 | 0.593827 | 0.60053 | 0.59398 |
|  | $E(10)$ | 0.8797 | 0.86128 | 0.8788 | 0.878804 | 0.884728 | 0.878825 |
|  | $E(12)$ | 1.2078 | 1.19335 | 1.20665 | 1.20685 | 1.20986 | 1.206755 |
|  | $E(14)$ | 1.5729 | 1.56711 | 1.57248 | 1.57275 | 1.57153 | 1.572609 |
|  | $E(16)$ | 1.9715 | 1.97646 | 1.97147 | 1.97166 | 1.9662 | 1.971543 |
|  | $E(18)$ | 2.3978 | 2.4142 | 2.39907 | 2.39906 | 2.39097 | 2.399039 |
|  | $E(20)$ | 2.85 | 2.871 | 2.85 | 2.8508 | 2.8434 | 2.850899 |
|  | $E(22)$ | 3.325 | 3.3358 | 3.323 | 3.323 | 3.3216 | 3.323246 |
|  | $E(24)$ | 3.812 | 3.7886 | 3.812 | 3.8126 | 3.8238 | 3.812526 |

**Table 1,** continued…

|  | $E(I)$ | Data | Expo1 $p = 3$ | Expo2 $p = 4$ | NS3 $p = 3$ | VMI $p = 2$ | LRV $p = 3$ |
|---|---|---|---|---|---|---|---|
| **232Th** | $E(2)$ | 0.049369 | 0.046986 | 0.051554 | 0.052111 | 0.051296 | 0.051566 |
|  | $E(4)$ | 0.16212 | 0.1537 | 0.16614 | 0.167346 | 0.167196 | 0.166222 |
|  | $E(6)$ | 0.3332 | 0.316576 | 0.3374 | 0.338902 | 0.340785 | 0.337598 |
|  | $E(8)$ | 0.5569 | 0.531959 | 0.5595 | 0.560796 | 0.564827 | 0.559809 |
|  | $E(10)$ | 0.827 | 0.796087 | 0.827 | 0.827747 | 0.833022 | 0.827454 |
|  | $E(12)$ | 1.1371 | 1.10507 | 1.1353 | 1.13508 | 1.14022 | 1.13562 |
|  | $E(14)$ | 1.4828 | 1.45483 | 1.4797 | 1.47864 | 1.48228 | 1.479878 |
|  | $E(16)$ | 1.8586 | 1.84111 | 1.8564 | 1.85473 | 1.85584 | 1.856284 |
|  | $E(18)$ | 2.2629 | 2.25936 | 2.2618 | 2.26004 | 2.25815 | 2.261381 |
|  | $E(20)$ | 2.6915 | 2.7047 | 2.6928 | 2.6916 | 2.687 | 2.692198 |
|  | $E(22)$ | 3.1442 | 3.1717 | 3.1467 | 3.1468 | 3.1403 | 3.146247 |
|  | $E(24)$ | 3.6196 | 3.6543 | 3.6216 | 3.6232 | 3.6166 | 3.621528 |
|  | $E(26)$ | 4.1162 | 4.1454 | 4.1159 | 4.1187 | 4.1144 | 4.116525 |
|  | $E(28)$ | 4.6318 | 4.6366 | 4.6291 | 4.6314 | 4.6325 | 4.630209 |
|  | $E(30)$ | 5.162 | 5.117 | 5.163 | 5.159 | 5.17 | 5.162035 |
| **234Th** | $E(2)$ | 0.04955 | 0.050087 | 0.049647 | 0.049329 | 0.049842 | 0.049435 |
|  | $E(4)$ | 0.163 | 0.163789 | 0.162964 | 0.162752 | 0.16383 | 0.162812 |
|  | $E(6)$ | 0.3365 | 0.336946 | 0.336318 | 0.336555 | 0.337332 | 0.336472 |
|  | $E(8)$ | 0.5648 | 0.564832 | 0.565063 | 0.565347 | 0.564863 | 0.565271 |
|  | $E(10)$ | 0.843 | 0.841828 | 0.842863 | 0.842452 | 0.841075 | 0.842586 |
|  | $E(12)$ | 1.1602 | 1.16077 | 1.16023 | 1.16036 | 1.16119 | 1.160311 |
| **230U** | $E(2)$ | 0.05172 | 0.051457 | 0.05271 | 0.052505 | 0.05199 | 0.052136 |
|  | $E(4)$ | 0.1695 | 0.167754 | 0.170521 | 0.170207 | 0.169995 | 0.169591 |
|  | $E(6)$ | 0.3471 | 0.344104 | 0.347429 | 0.347228 | 0.347769 | 0.346892 |
|  | $E(8)$ | 0.5782 | 0.57532 | 0.577644 | 0.577675 | 0.578488 | 0.578065 |
|  | $E(10)$ | 0.8564 | 0.855654 | 0.855554 | 0.855746 | 0.855978 | 0.856635 |
|  | $E(12)$ | 1.1757 | 1.17852 | 1.17568 | 1.17582 | 1.17505 | 1.175625 |
| **232U** | $E(2)$ | 0.047572 | 0.047695 | 0.048754 | 0.048399 | 0.048355 | 0.048638 |
|  | $E(4)$ | 0.15657 | 0.155803 | 0.158395 | 0.157746 | 0.158565 | 0.158166 |
|  | $E(6)$ | 0.3226 | 0.320368 | 0.32408 | 0.323469 | 0.325521 | 0.323842 |
|  | $E(8)$ | 0.541 | 0.537231 | 0.541061 | 0.540791 | 0.543394 | 0.540928 |
|  | $E(10)$ | 0.8058 | 0.80197 | 0.804667 | 0.804835 | 0.806704 | 0.804699 |
|  | $E(12)$ | 1.1115 | 1.10981 | 1.11027 | 1.11072 | 1.1107 | 1.110437 |
|  | $E(14)$ | 1.4537 | 1.4555 | 1.45326 | 1.45364 | 1.45139 | 1.453432 |
|  | $E(16)$ | 1.8281 | 1.8331 | 1.82896 | 1.82894 | 1.82546 | 1.828985 |
|  | $E(18)$ | 2.2315 | 2.23572 | 2.23257 | 2.23216 | 2.23011 | 2.232404 |
|  | $E(20)$ | 2.6597 | 2.6546 | 2.65898 | 2.65907 | 2.66301 | 2.659005 |

**Table 1,** continued…

|   | $E(I)$ | Data | Expo1 $p = 3$ | Expo2 $p = 4$ | NS3 $p = 3$ | VMI $p = 2$ | LRV $p = 3$ |
|---|---|---|---|---|---|---|---|
| **234U** | $E(2)$ | 0.043498 | 0.041039 | 0.045345 | 0.045495 | 0.044399 | 0.044599 |
|   | $E(4)$ | 0.143351 | 0.134995 | 0.146854 | 0.14715 | 0.145705 | 0.145124 |
|   | $E(6)$ | 0.296071 | 0.279609 | 0.299674 | 0.29998 | 0.299409 | 0.297353 |
|   | $E(8)$ | 0.49704 | 0.472491 | 0.499281 | 0.499432 | 0.500305 | 0.497103 |
|   | $E(10)$ | 0.7412 | 0.711086 | 0.741469 | 0.741344 | 0.743439 | 0.740218 |
|   | $E(12)$ | 1.0238 | 0.99263 | 1.02234 | 1.02191 | 1.02448 | 1.02258 |
|   | $E(14)$ | 1.3408 | 1.31408 | 1.33829 | 1.33763 | 1.33977 | 1.340101 |
|   | $E(16)$ | 1.6878 | 1.67202 | 1.68603 | 1.68533 | 1.68623 | 1.688727 |
|   | $E(18)$ | 2.063 | 2.06249 | 2.06253 | 2.06207 | 2.06129 | 2.064439 |
|   | $E(20)$ | 2.4642 | 2.48076 | 2.46511 | 2.46517 | 2.46277 | 2.463249 |
| **236U** | $E(2)$ | 0.045242 | 0.047315 | 0.046889 | 0.046234 | 0.047318 | 0.046404 |
|   | $E(4)$ | 0.149476 | 0.153943 | 0.152847 | 0.151405 | 0.155263 | 0.151654 |
|   | $E(6)$ | 0.309784 | 0.315445 | 0.313739 | 0.311887 | 0.318989 | 0.311999 |
|   | $E(8)$ | 0.52224 | 0.52748 | 0.525436 | 0.523718 | 0.532918 | 0.523553 |
|   | $E(10)$ | 0.7823 | 0.785804 | 0.783816 | 0.782695 | 0.791754 | 0.782294 |
|   | $E(12)$ | 1.0853 | 1.08627 | 1.08476 | 1.08447 | 1.09087 | 1.084063 |
|   | $E(14)$ | 1.4263 | 1.42481 | 1.42414 | 1.42465 | 1.42637 | 1.424565 |
|   | $E(16)$ | 1.8009 | 1.79747 | 1.79784 | 1.79886 | 1.79498 | 1.799369 |
|   | $E(18)$ | 2.2039 | 2.20035 | 2.20173 | 2.20281 | 2.19397 | 2.20391 |
|   | $E(20)$ | 2.6317 | 2.62965 | 2.6317 | 2.63235 | 2.62101 | 2.633484 |
|   | $E(22)$ | 3.0812 | 3.08167 | 3.0836 | 3.08355 | 3.07413 | 3.083253 |
|   | $E(24)$ | 3.55 | 3.55274 | 3.55328 | 3.55267 | 3.55165 | 3.548242 |
| **238U** | $E(2)$ | 0.044916 | 0.046941 | 0.046355 | 0.04572 | 0.047345 | 0.046329 |
|   | $E(4)$ | 0.14838 | 0.152766 | 0.151221 | 0.149839 | 0.155232 | 0.15116 |
|   | $E(6)$ | 0.30718 | 0.313113 | 0.310636 | 0.308893 | 0.318626 | 0.310549 |
|   | $E(8)$ | 0.5181 | 0.523718 | 0.520622 | 0.519061 | 0.531797 | 0.52053 |
|   | $E(10)$ | 0.7759 | 0.780406 | 0.77719 | 0.776266 | 0.789364 | 0.777114 |
|   | $E(12)$ | 1.0767 | 1.0791 | 1.07633 | 1.07627 | 1.08667 | 1.076287 |
|   | $E(14)$ | 1.4155 | 1.41579 | 1.41402 | 1.41475 | 1.41981 | 1.414014 |
|   | $E(16)$ | 1.7884 | 1.78658 | 1.78621 | 1.7874 | 1.78554 | 1.786238 |
|   | $E(18)$ | 2.1911 | 2.18764 | 2.18883 | 2.18998 | 2.18113 | 2.188878 |
|   | $E(20)$ | 2.6191 | 2.61522 | 2.61778 | 2.61837 | 2.60429 | 2.617829 |
|   | $E(22)$ | 3.0681 | 3.06564 | 3.06894 | 3.06865 | 3.05308 | 3.068964 |
|   | $E(24)$ | 3.5353 | 3.53531 | 3.53814 | 3.53708 | 3.52581 | 3.538134 |
|   | $E(26)$ | 4.0181 | 4.02069 | 4.0212 | 4.02019 | 4.02106 | 4.021166 |
|   | $E(28)$ | 4.517 | 4.51832 | 4.51387 | 4.51473 | 4.53753 | 4.513865 |

**Table 1**, continued…

|  | $E(I)$ | Data | Expo1 $p = 3$ | Expo2 $p = 4$ | NS3 $p = 3$ | VMI $p = 2$ | LRV $p = 3$ |
|---|---|---|---|---|---|---|---|
| **236Pu** | $E(2)$ | 0.04463 | 0.045477 | 0.045059 | 0.044751 | 0.04496 | 0.044898 |
| | $E(4)$ | 0.14745 | 0.148921 | 0.147994 | 0.147552 | 0.148339 | 0.14775 |
| | $E(6)$ | 0.3058 | 0.307014 | 0.3059 | 0.305695 | 0.306929 | 0.305765 |
| | $E(8)$ | 0.5157 | 0.516276 | 0.51551 | 0.515699 | 0.516682 | 0.515586 |
| | $E(10)$ | 0.7735 | 0.773025 | 0.773093 | 0.773454 | 0.773383 | 0.773287 |
| | $E(12)$ | 1.0743 | 1.07332 | 1.07431 | 1.07438 | 1.07305 | 1.074376 |
| | $E(14)$ | 1.4136 | 1.41285 | 1.414 | 1.41358 | 1.41211 | 1.41379 |
| | $E(16)$ | 1.786 | 1.78687 | 1.78584 | 1.786 | 1.78742 | 1.785903 |
| **238Pu** | $E(2)$ | 0.044076 | 0.045212 | 0.044883 | 0.044644 | 0.04496 | 0.044846 |
| | $E(4)$ | 0.145952 | 0.148251 | 0.147394 | 0.146892 | 0.148339 | 0.147312 |
| | $E(6)$ | 0.30338 | 0.306188 | 0.304836 | 0.304237 | 0.306929 | 0.304729 |
| | E(8) | 0.51358 | 0.516104 | 0.514459 | 0.513979 | 0.516682 | 0.514361 |
| | $E(10)$ | 0.77348 | 0.775095 | 0.773461 | 0.773264 | 0.773383 | 0.773401 |
| | $E(12)$ | 1.0801 | 1.08027 | 1.07898 | 1.07912 | 1.07305 | 1.078977 |
| | $E(14)$ | 1.4291 | 1.42873 | 1.4281 | 1.42849 | 1.41211 | 1.428144 |
| | $E(16)$ | 1.8185 | 1.81761 | 1.81782 | 1.81827 | 1.78742 | 1.817892 |
| | $E(18)$ | 2.2449 | 2.24402 | 2.24508 | 2.24537 | 2.19624 | 2.245142 |
| | $E(20)$ | 2.7057 | 2.70508 | 2.70672 | 2.70666 | 2.63621 | 2.706744 |
| | $E(22)$ | 3.1988 | 3.1979 | 3.1995 | 3.1991 | 3.10527 | 3.199484 |
| | $E(24)$ | 3.7208 | 3.7196 | 3.72014 | 3.7197 | 3.60161 | 3.720075 |
| | $E(26)$ | 4.2652 | 4.2673 | 4.2651 | 4.26552 | 4.12364 | 4.265164 |
| **240Pu** | $E(2)$ | 0.042824 | 0.043625 | 0.044081 | 0.044013 | 0.042812 | 0.044068 |
| | $E(4)$ | 0.14169 | 0.142947 | 0.14412 | 0.143965 | 0.141614 | 0.144096 |
| | $E(6)$ | 0.294319 | 0.295022 | 0.296843 | 0.296637 | 0.294043 | 0.296825 |
| | $E(8)$ | 0.49752 | 0.496929 | 0.49911 | 0.498903 | 0.496966 | 0.499112 |
| | $E(10)$ | 0.7478 | 0.745769 | 0.748 | 0.747737 | 0.746947 | 0.747923 |
| | $E(12)$ | 1.0418 | 1.03866 | 1.0403 | 1.04021 | 1.04058 | 1.040339 |
| | $E(14)$ | 1.3756 | 1.37275 | 1.3735 | 1.37351 | 1.37466 | 1.373553 |
| | $E(16)$ | 1.7456 | 1.74518 | 1.7448 | 1.74493 | 1.74628 | 1.744871 |
| | $E(18)$ | 2.152 | 2.15313 | 2.152 | 2.15185 | 2.15286 | 2.151713 |
| | $E(20)$ | 2.591 | 2.59377 | 2.592 | 2.59179 | 2.59208 | 2.591609 |
| | $E(22)$ | 3.061 | 3.0643 | 3.062 | 3.06236 | 3.0619 | 3.062205 |
| | $E(24)$ | 3.56 | 3.5619 | 3.56 | 3.56129 | 3.56052 | 3.561259 |
| | $E(26)$ | 4.088 | 4.08378 | 4.087 | 4.0864 | 4.08632 | 4.086639 |

**Table 1**, continued…

|  | $E(I)$ | Data | Expo1 $p = 3$ | Expo2 $p = 4$ | NS3 $p = 3$ | VMI $p = 2$ | LRV $p = 3$ |
|---|---|---|---|---|---|---|---|
| **242Pu** | $E(2)$ | 0.04454 | 0.046046 | 0.04536 | 0.044809 | 0.045684 | 0.045229 |
|  | $E(4)$ | 0.1473 | 0.150649 | 0.1488 | 0.14769 | 0.15072 | 0.148553 |
|  | $E(6)$ | 0.3064 | 0.310399 | 0.3075 | 0.306169 | 0.311837 | 0.307155 |
|  | $E(8)$ | 0.5181 | 0.521881 | 0.5183 | 0.517304 | 0.524908 | 0.518019 |
|  | $E(10)$ | 0.7786 | 0.781667 | 0.7781 | 0.777768 | 0.785642 | 0.777933 |
|  | $E(12)$ | 1.0844 | 1.08631 | 1.0834 | 1.08393 | 1.08999 | 1.083489 |
|  | $E(14)$ | 1.4317 | 1.43235 | 1.4308 | 1.43192 | 1.43431 | 1.431085 |
|  | $E(16)$ | 1.8167 | 1.81626 | 1.8166 | 1.81777 | 1.81541 | 1.816922 |
|  | $E(18)$ | 2.236 | 2.23452 | 2.237 | 2.2374 | 2.23052 | 2.237005 |
|  | $E(20)$ | 2.686 | 2.68351 | 2.687 | 2.68677 | 2.67723 | 2.687142 |
|  | $E(22)$ | 3.163 | 3.15956 | 3.163 | 3.1619 | 3.15345 | 3.162948 |
|  | $E(24)$ | 3.662 | 3.65891 | 3.66 | 3.65892 | 3.65734 | 3.659839 |
|  | $E(26)$ | 4.172 | 4.17768 | 4.173 | 4.17412 | 4.18729 | 4.173038 |
| **244Pu** | $E(2)$ | 0.0442 | 0.048448 | 0.04663 | 0.043968 | 0.050913 | 0.045509 |
|  | $E(4)$ | 0.155 | 0.158174 | 0.153157 | 0.14713 | 0.166258 | 0.15055 |
|  | $E(6)$ | 0.3179 | 0.325141 | 0.316661 | 0.308829 | 0.339602 | 0.313123 |
|  | $E(8)$ | 0.535 | 0.545226 | 0.533969 | 0.526881 | 0.564046 | 0.530513 |
|  | $E(10)$ | 0.8024 | 0.814204 | 0.801609 | 0.797669 | 0.833448 | 0.799293 |
|  | $E(12)$ | 1.1159 | 1.12772 | 1.11574 | 1.11634 | 1.14271 | 1.115321 |
|  | $E(14)$ | 1.471 | 1.48125 | 1.47208 | 1.47704 | 1.48768 | 1.473741 |
|  | $E(16)$ | 1.8635 | 1.87007 | 1.86577 | 1.87325 | 1.86496 | 1.868982 |
|  | $E(18)$ | 2.289 | 2.28915 | 2.29123 | 2.29806 | 2.27178 | 2.294761 |
|  | $E(20)$ | 2.742 | 2.73313 | 2.74195 | 2.74452 | 2.70581 | 2.74408 |
|  | $E(22)$ | 3.215 | 3.19612 | 3.21009 | 3.20586 | 3.16509 | 3.209226 |
|  | $E(24)$ | 3.69 | 3.67153 | 3.68599 | 3.67574 | 3.64795 | 3.681775 |
|  | $E(26)$ | 4.149 | 4.15177 | 4.15727 | 4.14842 | 4.15295 | 4.152586 |
|  | $E(28)$ | 4.61 | 4.62762 | 4.60724 | 4.61882 | 4.67881 | 4.611806 |

**Table 2.** The corresponding fits of our calculation are given in the units of [MeV].

| Nucleus | $A \times 10^{-3}$ | $a \times 10^{-5}$ | $B \times 10^{-7}$ |
|---|---|---|---|
| 224Th | 16.39883 | 73.42685 | $-152.1694$ |
| 226Th | 12.71619 | 38.96642 | $-58.41588$ |
| 228Th | 10.29599 | 23.01033 | $-25.83317$ |
| 230Th | 9.358425 | 14.61316 | $-8.382393$ |
| 232Th | 8.887517 | 15.043 | $-12.6447$ |
| 234Th | 8.306782 | 2.22513 | 38.58171 |
| 230U | 8.888611 | 9.574225 | 13.05427 |
| 232U | 8.304480 | 9.915433 | $-0.2278621$ |
| 234U | 7.610906 | 8.911491 | $-0.8635747$ |
| 236U | 7.882479 | 7.317457 | 3.542552 |
| 238U | 7.884489 | 8.132648 | 0.5960606 |
| 236Pu | 7.566771 | 3.742624 | 14.78418 |
| 238Pu | 7.581544 | 5.310031 | 1.783539 |
| 240Pu | 7.486860 | 7.198797 | $-2.938037$ |
| 242Pu | 7.644267 | 5.160483 | 5.100313 |
| 244Pu | 7.627230 | 1.564067 | 18.59270 |

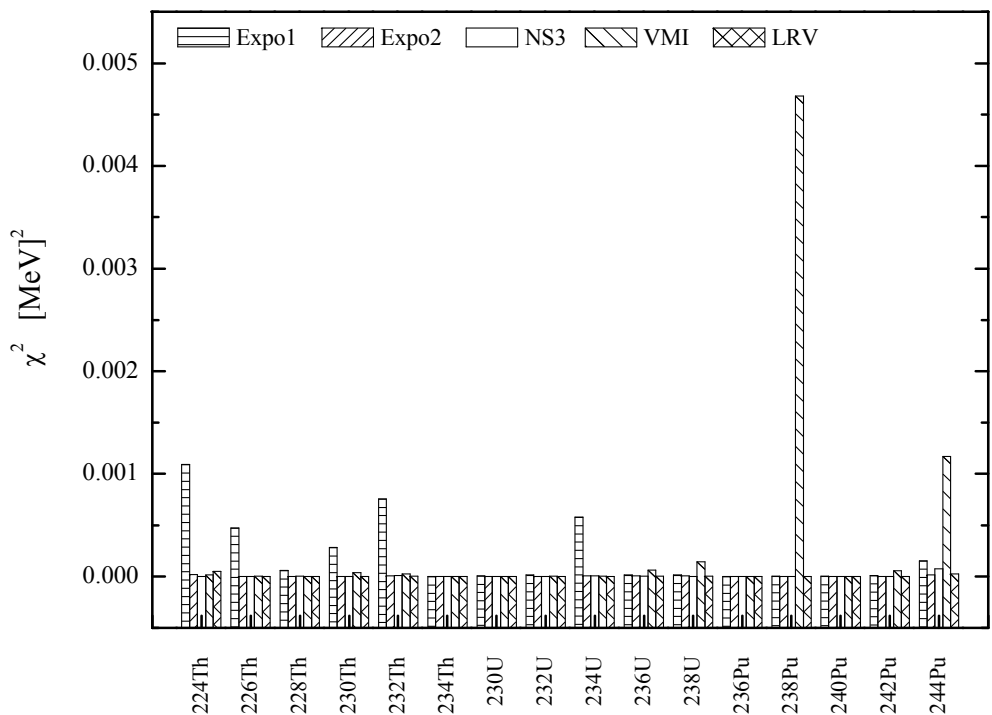

**Figure 1.** Deviation from data of various mentioned calculations.
(The zero level is shifted upward for convenience)

## Conclusion

In this work, a simple linear expression is deduced to calculate the energy levels of the ground state rotational band of deformed even-even Actinide nuclei. It includes three parameters *A*, *a*, and *B* which are determined straight forward using linear least squares fitting. The results of our calculation show good agreement with data in comparison with other calculations.

We observe that the correction parameter *B* bears small values and could be neglected in some cases without affecting the results. This is an indication for which the $\beta$- and $\gamma$-vibrations do not contribute much to the ground state band. On the other hand, the moment of inertia and its variation at high spin states show considerable contributions. Our simple expression can also be employed to study some nuclear effects in deformed even-even nuclei, including the backbending effects in some nuclei and possibly the variation of moment of inertia with high energy spin states.